\documentclass[runningheads]{llncs}
\usepackage{graphicx}
\makeatletter
\@ifundefined{lhs2tex.lhs2tex.sty.read}%
  {\@namedef{lhs2tex.lhs2tex.sty.read}{}%
   \newcommand\SkipToFmtEnd{}%
   \newcommand\EndFmtInput{}%
   \long\def\SkipToFmtEnd#1\EndFmtInput{}%
  }\SkipToFmtEnd

\newcommand\ReadOnlyOnce[1]{\@ifundefined{#1}{\@namedef{#1}{}}\SkipToFmtEnd}
\usepackage{amstext}
\usepackage{amssymb}
\usepackage{stmaryrd}
\DeclareFontFamily{OT1}{cmtex}{}
\DeclareFontShape{OT1}{cmtex}{m}{n}
  {<5><6><7><8>cmtex8
   <9>cmtex9
   <10><10.95><12><14.4><17.28><20.74><24.88>cmtex10}{}
\DeclareFontShape{OT1}{cmtex}{m}{it}
  {<-> ssub * cmtt/m/it}{}

\DeclareFontShape{OT1}{cmtt}{bx}{n}
  {<5><6><7><8>cmtt8
   <9>cmbtt9
   <10><10.95><12><14.4><17.28><20.74><24.88>cmbtt10}{}
\DeclareFontShape{OT1}{cmtex}{bx}{n}
  {<-> ssub * cmtt/bx/n}{}

\newcommand{\Conid}[1]{\mathit{#1}}
\newcommand{\Varid}[1]{\mathit{#1}}
\newcommand{\anonymous}{\kern0.06em \vbox{\hrule\@width.5em}}
\newcommand{\plus}{\mathbin{+\!\!\!+}}

\newcommand{\rbind}{\mathbin{=\mkern-6.7mu<\!\!\!<}}%

\usepackage{polytable}

\@ifundefined{mathindent}%
  {\newdimen\mathindent\mathindent\leftmargini}%
  {}%

\def\resethooks{%
  \global\let\SaveRestoreHook\empty
  \global\let\ColumnHook\empty}
\newcommand*{\savecolumns}[1][default]%
  {\g@addto@macro\SaveRestoreHook{\savecolumns[#1]}}
\newcommand*{\restorecolumns}[1][default]%
  {\g@addto@macro\SaveRestoreHook{\restorecolumns[#1]}}
\newcommand*{\aligncolumn}[2]%
  {\g@addto@macro\ColumnHook{\column{#1}{#2}}}

\resethooks

\newcommand{\onelinecommentchars}{\quad-{}- }
\newcommand{\commentbeginchars}{\enskip\{-}
\newcommand{\commentendchars}{-\}\enskip}

\newcommand{\visiblecomments}{%
  \let\onelinecomment=\onelinecommentchars
  \let\commentbegin=\commentbeginchars
  \let\commentend=\commentendchars}

\newcommand{\invisiblecomments}{%
  \let\onelinecomment=\empty
  \let\commentbegin=\empty
  \let\commentend=\empty}

\visiblecomments

\newlength{\blanklineskip}
\newcommand{\hsindent}[1]{\quad}%
\let\hspre\empty
\let\hspost\empty

\EndFmtInput
\makeatother
\ReadOnlyOnce{polycode.fmt}%
\makeatletter

\newcommand{\hsnewpar}[1]%
  {{\parskip=0pt\parindent=0pt\par\vskip #1\noindent}}

\newcommand{\hscodestyle}{}

\newcommand{\sethscode}[1]%
  {\expandafter\let\expandafter\hscode\csname #1\endcsname
   \expandafter\let\expandafter\endhscode\csname end#1\endcsname}

  {\par\noindent
   \advance\leftskip\mathindent
   \hscodestyle
   \let\\=\@normalcr
   \let\hspre\(\let\hspost\)%
   \pboxed}%
  {\endpboxed\)%
   \par\noindent
   \ignorespacesafterend}

  {\hsnewpar\abovedisplayskip
   \advance\leftskip\mathindent
   \hscodestyle
   \let\hspre\(\let\hspost\)%
   \pboxed}%
  {\endpboxed%
   \hsnewpar\belowdisplayskip
   \ignorespacesafterend}

  {\hsnewpar\abovedisplayskip
   \advance\leftskip\mathindent
   \hscodestyle
   \let\\=\@normalcr
   \(\pboxed}%
  {\endpboxed\)%
   \hsnewpar\belowdisplayskip
   \ignorespacesafterend}

\newcommand{\plainhs}{\sethscode{plainhscode}}

\plainhs

  {\hsnewpar\abovedisplayskip
   \advance\leftskip\mathindent
   \hscodestyle
   \let\\=\@normalcr
   \(\parray}%
  {\endparray\)%
   \hsnewpar\belowdisplayskip
   \ignorespacesafterend}

  {\parray}{\endparray}

  {\(\parray}{\endparray\)}

\def\codeframewidth{\arrayrulewidth}
\RequirePackage{calc}

  {\parskip=\abovedisplayskip\par\noindent
   \hscodestyle
   \arrayrulewidth=\codeframewidth
   \tabular{@{}|p{\linewidth-2\arraycolsep-2\arrayrulewidth-2pt}|@{}}%
   \hline\framedhslinecorrect\\{-1.5ex}%
   \let\endoflinesave=\\
   \let\\=\@normalcr
   \(\pboxed}%
  {\endpboxed\)%
   \framedhslinecorrect\endoflinesave{.5ex}\hline
   \endtabular
   \parskip=\belowdisplayskip\par\noindent
   \ignorespacesafterend}

\newcommand{\framedhslinecorrect}[2]%
  {#1[#2]}

  {\(\def\column##1##2{}%
   \let\>\undefined\let\<\undefined\let\\\undefined
   \newcommand\>[1][]{}\newcommand\<[1][]{}\newcommand\\[1][]{}%
   \def\fromto##1##2##3{##3}%
   }{\) }%

  {\let\orighscode=\hscode
   \let\origendhscode=\endhscode
   \def\endhscode{\def\hscode{\endgroup\def\@currenvir{hscode}\\}\begingroup}
   \orighscode\def\hscode{\endgroup\def\@currenvir{hscode}}}%
  {\origendhscode
   \global\let\hscode=\orighscode
   \global\let\endhscode=\origendhscode}%

\makeatother
\EndFmtInput
\usepackage[hyphens]{url}

\usepackage{relsize}
\usepackage{bussproofs}
\usepackage{mathtools}
\usepackage{bm}
\usepackage[autostyle]{csquotes}

\usepackage{stackengine}
\stackMath

\usepackage{xcolor}

\usepackage{float}
\floatstyle{boxed}
\restylefloat{figure}

\usepackage{syntax}

\usepackage{dashrule}

\newenvironment{dummy}{\color{gray}}{}

\newcommand{\deferred}[1]{}%
\newcommand{\integer}{\ensuremath{\mathbb{Z}}}
\newcommand{\readInput}[2]{\ensuremath{[\, \triangleright\, #1 \,]^{\,#2\,}}}
\newcommand{\writeOutputSimple}[1]{\ensuremath{[\,#1\, \triangleright\, ]}}
\newcommand{\restCh}[1]{\ensuremath{\mathop{\chAngle #1 \revChAngle}}}
\newcommand{\chAngle}{\raisebox{-1pt}{$\boldsymbol{\angle}$}}
\DeclareRobustCommand{\revChAngle}{\text{\reflectbox{\chAngle}}}
\newcommand{\loopExit}{\ensuremath{{\mathbf{E}}}}
\newcommand{\loopArr}{\ensuremath{{\rightarrow^\mathbf{E}}}}

\newcommand{\nat}{\ensuremath{\mathbb{N}}}

\let\Conid=\mathsf

\usepackage{listings}
\begin{document}
\title{A Framework for Generating Diverse Haskell-I/O Exercise Tasks}
\author{Oliver Westphal\orcidID{0000-0001-8947-0348}}
\authorrunning{O. Westphal}
\institute{University of Duisburg-Essen\\
\email{oliver.westphal@uni-due.de}}
\maketitle              %
\begin{abstract}
  We present the design of a framework to describe parametrized exercise tasks on Haskell-I/O programming.
  Parametrized tasks can be instantiated randomly to quickly generate different instances of a task.
  Such automatic task generation is useful in many different ways.
  Manual task creation can be a time-consuming process, so formulating a task design once and then automatically generating different variations can save valuable time for the educator.
  The descriptions of tasks also serve as easy to understand documentation and can be reused in new task designs.
  On the student's side automatic task generation, together with an automated assessment system, enables practicing on as many fresh exercise tasks as needed.
  Students can also each be given a slightly different version of tasks, reducing issues regarding plagiarism arising naturally in an e-learning environment.
  Our task generation is centered around a specification language for I/O behavior we developed in earlier work.
  The task generation framework, an embedded domain specific language in Haskell, provides powerful primitives for the creation of various artifacts from specifications, including program code.
  We do not go into detail on the technical realization of these primitives.
  Our focus is on showcasing how such artifacts can be used as an alternative to the verbal description of requirements for different types of programming exercise tasks.
  By doing so, we are able to automatically generate a diverse range of task designs.
\end{abstract}

\newcommand{\sampleSpec}{\ensuremath{\readInput{n}{\nat}(\readInput{x}{\integer}\restCh{\mathit{len}(x_A) = n_C}\loopExit)^\loopArr\writeOutputSimple{\mathit{sum}(x_A)}}}

\section{Introduction}
\label{sec:intro}
We have recently designed and implemented a language for specifying console I/O programs~\cite{TFPIE19-paper,flops20-paper} allowing us to formulate desired I/O behavior.
The I/O behavior of programs, written in Haskell, can be tested probabilistically against specified behavior.
We built this language to bring our testing capabilities of tasks on Haskell I/O more in line with how one can test tasks on pure programs, for example, using QuickCheck~\cite{claessen2000quickcheck}.
These automatic testing capabilities are used in the e-learning system \cite{rahn2008leipzig,DBLP:conf/abp/Waldmann17} we use in our course on programming paradigms~\cite{DBLP:conf/abp/SiegburgVW19}.

We now also aim to automatically generate the tasks themselves.
This has a variety of advantages.
Automatic task generation can help educators to create different variations of a common exercise task idea much quicker.
When combined with automated assessment students have the opportunity to practice with as many fresh exercise tasks as they need.
Automatically generated tasks can also be used to reduce plagiarism issues by giving students slight variations of the same task.
We will (only) present automatic generation for tasks on Haskell I/O in this work, since our specification language is designed to describe the I/O behavior of programs.
However, we believe our approach can be adapted for other types of exercise tasks as well.

Hand-written (programming) exercise tasks usually rely heavily on verbal descriptions.
For example we might pose a task like this:
\begin{flushleft}
\textit{%
``Read a positive integer $n$ from the console. Then, read $n$ integers one after the other. Finally, output their sum.''
}
\end{flushleft}
Such verbal descriptions are a big problem when trying to generate tasks automatically, as natural language generation is not exactly easy.
Because of this, many task generation systems rely on templates defining a fixed (verbal) frame for a task.
Such templates contain gaps to be filled to form a concrete task.
Different (randomized) choices to fill these gaps result in different task variations.
Depending on the domain for which tasks are generated, writing such a fixed framing, can be difficult.
For example, it is easy for many different math tasks, where the verbal frame can be something like ``Solve for $x$.'' together with a randomized equation.
For programming tasks finding a fixed and general verbal frame is more difficult.
Take, for example, the verbal description from above.
We could use a fixed verbal skeleton like ``First ... Then ... Finally'' and fill it with random predefined descriptions.
However, this is not a very flexible approach.
Instead, our approach uses artifacts like program code or example runs of a program to achieve a fixed descriptions.
Take, for example, the following task:
\label{page:ex-task}
\begin{hscode}\SaveRestoreHook
\column{B}{@{}>{\hspre}l<{\hspost}@{}}%
\column{3}{@{}>{\hspre}l<{\hspost}@{}}%
\column{9}{@{}>{\hspre}l<{\hspost}@{}}%
\column{20}{@{}>{\hspre}l<{\hspost}@{}}%
\column{22}{@{}>{\hspre}l<{\hspost}@{}}%
\column{E}{@{}>{\hspre}l<{\hspost}@{}}%
\>[B]{}\texttt{Give the programs interaction trace for input sequence 2, 4, 9.}{}\<[E]%
\\
\>[B]{}\Varid{prog}\mathbin{::}\Conid{IO}\;(){}\<[E]%
\\
\>[B]{}\Varid{prog}\mathrel{=}\mathbf{do}{}\<[E]%
\\
\>[B]{}\hsindent{3}{}\<[3]%
\>[3]{}\Varid{n}\leftarrow \Varid{readLn}{}\<[E]%
\\
\>[B]{}\hsindent{3}{}\<[3]%
\>[3]{}\mathbf{let}\;\Varid{loop}\;\Varid{s}\;\Varid{l}\mathrel{=}{}\<[E]%
\\
\>[3]{}\hsindent{6}{}\<[9]%
\>[9]{}\mathbf{if}\;\Varid{l}\mathrel{==}\Varid{n}\;{}\<[20]%
\>[20]{}\mathbf{then}\;\Varid{print}\;\Varid{s}{}\<[E]%
\\
\>[20]{}\mathbf{else}\;\mathbf{do}{}\<[E]%
\\
\>[20]{}\hsindent{2}{}\<[22]%
\>[22]{}\Varid{v}\leftarrow \Varid{readLn}{}\<[E]%
\\
\>[20]{}\hsindent{2}{}\<[22]%
\>[22]{}\Varid{loop}\;(\Varid{s}\mathbin{+}\Varid{v})\;(\Varid{l}\mathbin{+}\mathrm{1}){}\<[E]%
\\
\>[B]{}\hsindent{3}{}\<[3]%
\>[3]{}\Varid{loop}\;\mathrm{0}\;\mathrm{0}{}\<[E]%
\ColumnHook
\end{hscode}\resethooks
The verbal description does not need to change, apart from maybe the given input sequence, no matter what we give as the program text.
However, a solution to this task is fundamentally different than in the previous example.
Instead of requiring a correctly behaving program, we simply ask for one specific run of a given program.
Executing programs ``by hand'' is an important skill to have, but this task is not a substitute for a more open-ended programming task.

Our main contribution is to show that there is in fact a rich spectrum of different I/O tasks between these two extremes, as we will explore in section~\ref{sec:examples}.
Moving along this spectrum yields a diverse collection of tasks requiring different skills to solve.
We design tasks ranging from program reading and comprehension over completion of partial programs all the way to writing original programs.

These designs are expressed in a newly developed Haskell embedded domain specific language (EDSL).
This EDSL has two purposes.
On the one hand it provides components to describe task designs and generate concrete task instances thereof.
On the other hand, and equally important, it encourages descriptions that separate orthogonal aspects of task designs.
This separation makes maintaining, expanding and reusing task designs much easier for the educator.
Especially, when working with task designs created by another author, as it can also act as a form of documentation.

We build on top of the existing implementation of the specification language from our previous work.
Starting from a specification, hand-written or generated randomly, we derive example runs and programs satisfying the specification.
These artifacts are then used to build tasks.
For example, we can give examples runs and ask for a corresponding program.
Or we give a program and ask for runs of that program.
The testing capabilities of the specification language allow us to automatically check solution candidates for such generated tasks for correctness.

We will not go into the technical details of how we create these artifacts in the implementation.
Instead this presentation focuses on the framework's versatility in expressing interesting task ideas and generating variations.

We will first give a short overview of the previously introduced specification language.
Next we will define the EDSL for describing tasks.
Using both the specification language and the language of tasks together, we show how to design a diverse range of exercise tasks on Haskell-I/O.

\section{Specifying and testing I/O behavior}
\label{sec:specs}
Our previous work~\cite{TFPIE19-paper} introduced a specification language for I/O behavior.
The goal of the language is to enable easy QuickCheck testing for I/O behavior.
Specifications expressed in the language describe program behavior in terms of traces, i.e., sequences of read and written values, a program should produce.
Programs are tested against specifications by repeatedly checking for different (carefully randomized) inputs whether the trace of that program matches the specification.
We will not present a formal introduction of specifications as we did in our previous work~\cite{TFPIE19-paper}.
A high-level overview of the language's features is enough for this work.

The language has typical elements of a standard imperative language, but varies in some aspects, most notably the use of variables.
It defines primitives for reading and writing values from and to the console, a branching construct to choose sub-specifications based on Boolean conditions, and an iteration construct.
Iteration is done through loops, but with explicit exit markers instead of a termination condition.
Here is the summation behavior from the previous section expressed as a specification:
$$ \sampleSpec $$
Reading and writing, i.e., the primitive I/O action we want to observe, are written as $\readInput{x}{\tau}$ and $\writeOutputSimple{t}$, with $x$ specifying the variable into which to read the value and $t$ being a term describing the value to be written.
Reading actions also have an annotation $\tau$ specifying the set of legal inputs we expect at that point.
For example, we expect the first value read in the example specification to be a natural number.
Branching on a Boolean condition $c$ is written as $s_F \restCh{c} s_T$, choosing the right branch $s_T$ if $c$ holds. Sub-specifications to be repeated are marked by ${}^\loopArr$, and such a loop terminates on reaching an exit marker $\loopExit$.

Variables in the specification language behave differently compared to classical imperative languages.
They accumulate all values read into them.
Variables are then either used as single values, i.e., the last value read into them, or as a list of all past values of that variable.
The subscript of a variable indicates how it is used.
\textit{C} stands for the \underline{c}urrent value and \textit{A} for \underline{a}ll values.
By design, specifications only define how inputs and outputs are interleaved and what the output values should be.
They cannot describe internal states of a program.

The implementation, accompanying our previous work, exposes, besides constructors for \ensuremath{\Conid{Specification}}s, a simple API for testing programs against specifications.
Testing relies on programs being expressed in a variant of the standard Haskell \ensuremath{\Conid{IO}} monad in which we can observe I/O effects~\cite{TFPIE19-paper,swierstra2007}.
This allows us to take a program, i.e., a value of the inspectable \ensuremath{\mathit{IO}_\mathit{rep}} type, and run it on an input sequence.
\begin{hscode}\SaveRestoreHook
\column{B}{@{}>{\hspre}l<{\hspost}@{}}%
\column{E}{@{}>{\hspre}l<{\hspost}@{}}%
\>[B]{}\Varid{runProgram}\mathbin{::}[\mskip1.5mu \Conid{String}\mskip1.5mu]\to \mathit{IO}_\mathit{rep}\;()\to \Conid{Trace}{}\<[E]%
\ColumnHook
\end{hscode}\resethooks
We can then check whether the program run, encoded by its \ensuremath{\Conid{Trace}} of I/O actions, satisfies the behavior described by some \ensuremath{\Conid{Specification}}.
\begin{hscode}\SaveRestoreHook
\column{B}{@{}>{\hspre}l<{\hspost}@{}}%
\column{E}{@{}>{\hspre}l<{\hspost}@{}}%
\>[B]{}\Varid{accept}\mathbin{::}\Conid{Specification}\to \Conid{Trace}\to \Conid{Bool}{}\<[E]%
\ColumnHook
\end{hscode}\resethooks
Repeating this process for randomly generated input sequences, we formulate a QuickCheck \ensuremath{\Conid{Property}} stating that a program satisfies a specification.
\begin{hscode}\SaveRestoreHook
\column{B}{@{}>{\hspre}l<{\hspost}@{}}%
\column{E}{@{}>{\hspre}l<{\hspost}@{}}%
\>[B]{}\Varid{fulfills}\mathbin{::}\mathit{IO}_\mathit{rep}\;()\to \Conid{Specification}\to \Conid{Property}{}\<[E]%
\ColumnHook
\end{hscode}\resethooks
We also provide an interpreter to turn specifications into executable programs.
\begin{hscode}\SaveRestoreHook
\column{B}{@{}>{\hspre}l<{\hspost}@{}}%
\column{E}{@{}>{\hspre}l<{\hspost}@{}}%
\>[B]{}\Varid{buildComputation}\mathbin{::}\Conid{Specification}\to \mathit{IO}_\mathit{rep}\;(){}\<[E]%
\ColumnHook
\end{hscode}\resethooks
See~\cite{flops20-paper} for details on this.

\section{Describing parameterized tasks}
\label{sec:tasks}
This section introduces a small embedded domain specific language in Haskell to describe exercise tasks, including automatically checkable requirements.
The language enables clear and concise descriptions of parameterized tasks.
Descriptions can be built from orthogonal components allowing for quick and easy reuse and modification.
We will use this language in section~\ref{sec:examples} to discuss different categories of exercise tasks on Haskell I/O.
The EDSL itself, however, can be used to describe parameterized tasks on any topic.

The language consists of three separate components.
Descriptions of concrete exercise tasks, called task instances, a (sub-)language for describing requirements of correct solutions, and ways to express general task designs, i.e., generators for concrete tasks.
Generally speaking, task designs bundle up generators for parameters together with a recipe for turning parameters into task instances.

The design goals of the EDSL are as follows:
\begin{itemize}
  \item Clearly and concisely communicate the task's idea through its description, \emph{without} exposing computational details or requiring knowledge thereof.
  \item Separate the basic building blocks of tasks into orthogonal and reusable components.
  \item The main purpose of a task's description is to be read by educators.
        Automatically checking whether a solution candidate fulfills a task's requirements is only a secondary feature. %
\end{itemize}

First off we need a data type for concrete task instances.
\begin{hscode}\SaveRestoreHook
\column{B}{@{}>{\hspre}l<{\hspost}@{}}%
\column{3}{@{}>{\hspre}l<{\hspost}@{}}%
\column{E}{@{}>{\hspre}l<{\hspost}@{}}%
\>[B]{}\mathbf{data}\;\Conid{TaskInstance}\;\Varid{s}\mathrel{=}\Conid{TaskInstance}{}\<[E]%
\\
\>[B]{}\hsindent{3}{}\<[3]%
\>[3]{}\{\mskip1.5mu \Varid{question}\mathbin{::}\Conid{Description}{}\<[E]%
\\
\>[B]{}\hsindent{3}{}\<[3]%
\>[3]{},\Varid{given}\mathbin{::}\Conid{Maybe}\;\Varid{s}{}\<[E]%
\\
\>[B]{}\hsindent{3}{}\<[3]%
\>[3]{},\Varid{requires}\mathbin{::}\Conid{Require}\;\Varid{s}\mskip1.5mu\}{}\<[E]%
\ColumnHook
\end{hscode}\resethooks
The type parameter \ensuremath{\Varid{s}} represents the type of solution the \ensuremath{\Conid{TaskInstance}} expects.
For simplicity we treat \ensuremath{\Conid{Description}} as an abstract string-like type for which we assume standard layout combinators exist~\cite{hughes1995}.
Each \ensuremath{\Conid{TaskInstance}} can have a default \ensuremath{\Varid{given}} value of type \ensuremath{\Varid{s}}.
By convention we treat this value as a somehow incomplete version of a correct solution to be used as a starting point for solving the task.

The \ensuremath{\Conid{Require}} type encodes the conditions under which a solution candidate is deemed correct.
Requirements are not constructed directly, instead the EDSL provides constructor functions for different requirements.
The simplest requirements are predicates on the solution type \ensuremath{\Varid{s}}.
\begin{hscode}\SaveRestoreHook
\column{B}{@{}>{\hspre}l<{\hspost}@{}}%
\column{E}{@{}>{\hspre}l<{\hspost}@{}}%
\>[B]{}\Varid{requirePure}\mathbin{::}(\Varid{s}\to \Conid{Bool})\to \Conid{Require}\;\Varid{s}{}\<[E]%
\ColumnHook
\end{hscode}\resethooks
For more complex requirements we use QuickCheck's \ensuremath{\Conid{Property}} type to enable randomized testing.
QuickCheck also provides feedback in case the \ensuremath{\Conid{Property}} fails.
\begin{hscode}\SaveRestoreHook
\column{B}{@{}>{\hspre}l<{\hspost}@{}}%
\column{E}{@{}>{\hspre}l<{\hspost}@{}}%
\>[B]{}\Varid{requireProp}\mathbin{::}(\Varid{s}\to \Conid{Property})\to \Conid{Require}\;\Varid{s}{}\<[E]%
\ColumnHook
\end{hscode}\resethooks
We can also add an arbitrary \ensuremath{\Conid{IO}} pre-processing step to a requirement.
\begin{hscode}\SaveRestoreHook
\column{B}{@{}>{\hspre}l<{\hspost}@{}}%
\column{E}{@{}>{\hspre}l<{\hspost}@{}}%
\>[B]{}\Varid{after}\mathbin{::}\Conid{Require}\;\Varid{s'}\to (\Varid{s}\to \Conid{IO}\;(\Conid{Maybe}\;\Varid{s'}))\to \Conid{Require}\;\Varid{s}{}\<[E]%
\ColumnHook
\end{hscode}\resethooks
\ensuremath{\Conid{Maybe}} here indicates that pre-processing might fail, in which case the requirement is not fulfilled.
One usage of this combinator, we will see later, is to compile programs given as textual input to actual Haskell values usable in a \ensuremath{\Conid{Property}}.

We define a primitive for building the conjunction of two requirements.
\begin{hscode}\SaveRestoreHook
\column{B}{@{}>{\hspre}l<{\hspost}@{}}%
\column{E}{@{}>{\hspre}l<{\hspost}@{}}%
\>[B]{}(\mathbin{/\char92 })\mathbin{::}\Conid{Require}\;\Varid{s}\to \Conid{Require}\;\Varid{s}\to \Conid{Require}\;\Varid{s}{}\<[E]%
\ColumnHook
\end{hscode}\resethooks

Lastly we might require a correct solution to ``match'' the \ensuremath{\Varid{given}} value of the \ensuremath{\Conid{TaskInstance}}.
For example, filling in gaps in a given skeleton.
We define a class to specify what matching a skeleton means for a specific type.
\begin{hscode}\SaveRestoreHook
\column{B}{@{}>{\hspre}l<{\hspost}@{}}%
\column{3}{@{}>{\hspre}l<{\hspost}@{}}%
\column{E}{@{}>{\hspre}l<{\hspost}@{}}%
\>[B]{}\mathbf{class}\;\Conid{Matches}\;\Varid{s}\;\mathbf{where}{}\<[E]%
\\
\>[B]{}\hsindent{3}{}\<[3]%
\>[3]{}\Varid{matches}\mathbin{::}\Varid{s}\to \Varid{s}\to \Conid{Bool}{}\<[E]%
\ColumnHook
\end{hscode}\resethooks
Conceptually an instance of \ensuremath{\Conid{Matches}} defines a partial order on \ensuremath{\Varid{s}} where \ensuremath{\Varid{matches}\;\Varid{t}\;\Varid{s}} evaluates to \ensuremath{\Conid{True}} iff \ensuremath{\Varid{s}} is an extended version of the partial solution \ensuremath{\Varid{t}}.
\begin{hscode}\SaveRestoreHook
\column{B}{@{}>{\hspre}l<{\hspost}@{}}%
\column{E}{@{}>{\hspre}l<{\hspost}@{}}%
\>[B]{}\Varid{mustMatch}\mathbin{::}\Conid{Matches}\;\Varid{s}\Rightarrow \Varid{s}\to \Conid{Require}\;\Varid{s}{}\<[E]%
\\
\>[B]{}\Varid{mustMatch}\mathrel{=}\Varid{requirePure}\mathbin{\circ}\Varid{matches}{}\<[E]%
\ColumnHook
\end{hscode}\resethooks

Checking whether a requirement holds for some value will in general require \ensuremath{\Conid{IO}}.
Either to run QuickCheck or because we used \ensuremath{\Varid{after}}.
\begin{hscode}\SaveRestoreHook
\column{B}{@{}>{\hspre}l<{\hspost}@{}}%
\column{E}{@{}>{\hspre}l<{\hspost}@{}}%
\>[B]{}\Varid{check}\mathbin{::}\Conid{Require}\;\Varid{s}\to \Varid{s}\to \Conid{IO}\;\Conid{Bool}{}\<[E]%
\ColumnHook
\end{hscode}\resethooks

Being able to represent concrete tasks, we can now define parameterized tasks as regular Haskell functions from parameters to \ensuremath{\Conid{TaskInstance}} values.
For example, we can define a simple (non-I/O) task requiring adding up two numbers:
\begin{hscode}\SaveRestoreHook
\column{B}{@{}>{\hspre}l<{\hspost}@{}}%
\column{3}{@{}>{\hspre}l<{\hspost}@{}}%
\column{E}{@{}>{\hspre}l<{\hspost}@{}}%
\>[B]{}\Varid{taskAdd}\mathbin{::}\Conid{Int}\to \Conid{Int}\to \Conid{TaskInstance}\;\Conid{Int}{}\<[E]%
\\
\>[B]{}\Varid{taskAdd}\;\Varid{x}\;\Varid{y}\mathrel{=}\Conid{TaskInstance}{}\<[E]%
\\
\>[B]{}\hsindent{3}{}\<[3]%
\>[3]{}\{\mskip1.5mu \Varid{question}\mathrel{=}\Varid{text}\;(\text{\ttfamily \char34 Give~the~sum~of~\char34}\plus \Varid{show}\;\Varid{x}\plus \text{\ttfamily \char34 ~and~\char34}\plus \Varid{show}\;\Varid{y}){}\<[E]%
\\
\>[B]{}\hsindent{3}{}\<[3]%
\>[3]{},\Varid{given}\mathrel{=}\Conid{Nothing}{}\<[E]%
\\
\>[B]{}\hsindent{3}{}\<[3]%
\>[3]{},\Varid{requires}\mathrel{=}\Varid{exactAnswer}\;(\Varid{x}\mathbin{+}\Varid{y})\mskip1.5mu\}{}\<[E]%
\\[\blanklineskip]%
\>[B]{}\Varid{exactAnswer}\mathbin{::}(\Conid{Eq}\;\Varid{a},\Conid{Show}\;\Varid{a})\Rightarrow \Varid{a}\to \Conid{Require}\;\Varid{a}{}\<[E]%
\\
\>[B]{}\Varid{exactAnswer}\;\Varid{x}\mathrel{=}\Varid{requireProp}\mathbin{\$}\lambda \Varid{s}\to \Varid{s}\mathbin{===}\Varid{x}{}\<[E]%
\ColumnHook
\end{hscode}\resethooks
Defining \ensuremath{\Varid{exactAnswer}} in terms of QuickCheck's \ensuremath{(\mathbin{===})} operator, we get informative feedback from QuickCheck's output in case of a test failure.
Giving the wrong solution to an instance of the above task might, for example, result in the following error:
\begin{hscode}\SaveRestoreHook
\column{B}{@{}>{\hspre}l<{\hspost}@{}}%
\column{3}{@{}>{\hspre}l<{\hspost}@{}}%
\column{E}{@{}>{\hspre}l<{\hspost}@{}}%
\>[3]{}\texttt{>}\Varid{check}\;(\Varid{requires}\;(\Varid{taskAdd}\;\mathrm{2}\;\mathrm{3}))\;\mathrm{4}{}\<[E]%
\\
\>[3]{}\texttt{*** Failed! Falsified (after 1 test):}{}\<[E]%
\\
\>[3]{}\texttt{4 /= 5}{}\<[E]%
\ColumnHook
\end{hscode}\resethooks
The last step to automatic task generation is to couple a parameterized \ensuremath{\Conid{TaskInstance}} with a generator of its expected parameter.
\begin{hscode}\SaveRestoreHook
\column{B}{@{}>{\hspre}l<{\hspost}@{}}%
\column{3}{@{}>{\hspre}l<{\hspost}@{}}%
\column{E}{@{}>{\hspre}l<{\hspost}@{}}%
\>[B]{}\mathbf{data}\;\Conid{TaskDesign}\;\Varid{s}\mathrel{=}\forall \Varid{p}.\Conid{TaskDesign}{}\<[E]%
\\
\>[B]{}\hsindent{3}{}\<[3]%
\>[3]{}\{\mskip1.5mu \Varid{parameter}\mathbin{::}\Conid{Gen}\;\Varid{p}{}\<[E]%
\\
\>[B]{}\hsindent{3}{}\<[3]%
\>[3]{},\Varid{instantiate}\mathbin{::}\Varid{p}\to \Conid{TaskInstance}\;\Varid{s}\mskip1.5mu\}{}\<[E]%
\ColumnHook
\end{hscode}\resethooks
To instantiate a design we generate a parameter value and pass it to \ensuremath{\Varid{instantiate}}:
\begin{hscode}\SaveRestoreHook
\column{B}{@{}>{\hspre}l<{\hspost}@{}}%
\column{3}{@{}>{\hspre}l<{\hspost}@{}}%
\column{E}{@{}>{\hspre}l<{\hspost}@{}}%
\>[B]{}\Varid{generateTaskInstance}\mathbin{::}\Conid{TaskDesign}\;\Varid{s}\to \Conid{IO}\;(\Conid{TaskInstance}\;\Varid{s}){}\<[E]%
\\
\>[B]{}\Varid{generateTaskInstance}\;(\Conid{TaskDesign}\;\Varid{param}\;\Varid{inst})\mathrel{=}{}\<[E]%
\\
\>[B]{}\hsindent{3}{}\<[3]%
\>[3]{}\Varid{generate}\;(\Varid{inst}\mathbin{<\hspace{-1.6pt}\mathclap{\raisebox{0.1pt}{\scalebox{.8}{\$}}}\hspace{-1.6pt}>}\Varid{param}){}\<[E]%
\ColumnHook
\end{hscode}\resethooks
We define combinators to aid in our goal of clearly communicating both a task's idea and requirements.
Instead of using the \ensuremath{\Conid{TaskDesign}} constructor itself we use
\begin{hscode}\SaveRestoreHook
\column{B}{@{}>{\hspre}l<{\hspost}@{}}%
\column{E}{@{}>{\hspre}l<{\hspost}@{}}%
\>[B]{}\Varid{for}\mathbin{::}\Conid{Gen}\;\Varid{p}\to (\Varid{p}\to \Conid{TaskInstance}\;\Varid{s})\to \Conid{TaskDesign}\;\Varid{s}{}\<[E]%
\\
\>[B]{}\Varid{for}\mathrel{=}\Conid{TaskDesign}{}\<[E]%
\ColumnHook
\end{hscode}\resethooks
resulting in the general pattern
\begin{hscode}\SaveRestoreHook
\column{B}{@{}>{\hspre}l<{\hspost}@{}}%
\column{3}{@{}>{\hspre}l<{\hspost}@{}}%
\column{10}{@{}>{\hspre}l<{\hspost}@{}}%
\column{E}{@{}>{\hspre}l<{\hspost}@{}}%
\>[B]{}\Varid{someTask}\mathbin{::}\Conid{TaskDesign}\;\Varid{s}{}\<[E]%
\\
\>[B]{}\Varid{someTask}\mathrel{=}\Varid{for}\;\Varid{someRandomParameter}\;\Varid{doSomething}{}\<[E]%
\\
\>[B]{}\hsindent{3}{}\<[3]%
\>[3]{}\mathbf{where}\;{}\<[10]%
\>[10]{}\Varid{someRandomPrameter}\mathbin{::}\Conid{Gen}\;\Varid{p}{}\<[E]%
\\
\>[10]{}\Varid{doSomthing}\mathbin{::}\Varid{p}\to \Conid{TaskInstance}\;\Varid{s}{}\<[E]%
\ColumnHook
\end{hscode}\resethooks
The arguments to \ensuremath{\Varid{for}} are deliberately named to make the expression read like a high-level description of the task.

Generators can be combined and modified by specialized instantiations of well known combinators on monads and arrows~\cite{hughes2000}.
The new names of these combinators reflect their domain specific usage and thereby aid in hiding computational details from task descriptions.
The first of these combinators is
\begin{hscode}\SaveRestoreHook
\column{B}{@{}>{\hspre}l<{\hspost}@{}}%
\column{E}{@{}>{\hspre}l<{\hspost}@{}}%
\>[B]{}\Varid{fixed}\mathbin{::}\Varid{p}\to \Conid{Gen}\;\Varid{p}{}\<[E]%
\\
\>[B]{}\Varid{fixed}\mathrel{=}\Varid{pure}{}\<[E]%
\ColumnHook
\end{hscode}\resethooks
allowing us to write \ensuremath{\Varid{for}\;(\Varid{fixed}\;\Varid{parameter})\;\Varid{doSomething}} when we do not want to generate randomized tasks but still use the EDSL to communicate our design.
Next we define
\begin{hscode}\SaveRestoreHook
\column{B}{@{}>{\hspre}l<{\hspost}@{}}%
\column{E}{@{}>{\hspre}l<{\hspost}@{}}%
\>[B]{}\Varid{from}\mathbin{::}(\Varid{a}\to \Conid{Gen}\;\Varid{b})\to \Conid{Gen}\;\Varid{a}\to \Conid{Gen}\;\Varid{b}{}\<[E]%
\\
\>[B]{}\Varid{from}\mathrel{=}(\rbind ){}\<[E]%
\ColumnHook
\end{hscode}\resethooks
so that we can describe parameter generators in terms of existing generators:
\begin{hscode}\SaveRestoreHook
\column{B}{@{}>{\hspre}l<{\hspost}@{}}%
\column{3}{@{}>{\hspre}l<{\hspost}@{}}%
\column{E}{@{}>{\hspre}l<{\hspost}@{}}%
\>[3]{}\Varid{for}\;(\Varid{randomParameterB}\mathbin{`\Varid{from}`}\Varid{randomParameterA})\;\Varid{doSomething}{}\<[E]%
\ColumnHook
\end{hscode}\resethooks
Splitting up parameter generators separates different layers of randomness.
Each layer can then easily be changed independently.
For example, we can change \ensuremath{\Varid{randomParameterA}} to a fixed example
\begin{hscode}\SaveRestoreHook
\column{B}{@{}>{\hspre}l<{\hspost}@{}}%
\column{3}{@{}>{\hspre}l<{\hspost}@{}}%
\column{E}{@{}>{\hspre}l<{\hspost}@{}}%
\>[3]{}\Varid{for}\;(\Varid{randomParameterB}\mathbin{`\Varid{from}`}\Varid{fixed}\;\Varid{a})\;\Varid{doSomething}{}\<[E]%
\ColumnHook
\end{hscode}\resethooks
while \ensuremath{\Varid{randomParameterB}} is untouched.
The parameter used to create a task instance is still randomized, but with one layer of randomness less.

The \ensuremath{\Varid{instantiate}} function can only ever take a single argument.
For tasks with multiple randomized parameters we define combinators for tuple generators.
\begin{hscode}\SaveRestoreHook
\column{B}{@{}>{\hspre}l<{\hspost}@{}}%
\column{E}{@{}>{\hspre}l<{\hspost}@{}}%
\>[B]{}(\mathrel{\&\hspace{-2pt}\&\hspace{-2pt}\&\hspace{-2pt}\&})\mathbin{::}\Conid{Monad}\;\Varid{m}\Rightarrow (\Varid{a}\to \Varid{m}\;\Varid{b})\to (\Varid{a}\to \Varid{m}\;\Varid{b'})\to \Varid{a}\to \Varid{m}\;(\Varid{b},\Varid{b'}){}\<[E]%
\\
\>[B]{}(\mathrel{*\hspace{-3pt}*\hspace{-3pt}*\hspace{-1pt}*})\mathbin{::}\Conid{Monad}\;\Varid{m}\Rightarrow (\Varid{a}\to \Varid{m}\;\Varid{b})\to (\Varid{a'}\to \Varid{m}\;\Varid{b'})\to (\Varid{a},\Varid{a'})\to \Varid{m}\;(\Varid{b},\Varid{b'}){}\<[E]%
\ColumnHook
\end{hscode}\resethooks
Looking ahead to section~\ref{sec:examples}, here is an example of how these combinators can be used to describe a \ensuremath{\Conid{TaskDesign}}.
\begin{hscode}\SaveRestoreHook
\column{B}{@{}>{\hspre}l<{\hspost}@{}}%
\column{3}{@{}>{\hspre}l<{\hspost}@{}}%
\column{5}{@{}>{\hspre}l<{\hspost}@{}}%
\column{E}{@{}>{\hspre}l<{\hspost}@{}}%
\>[3]{}\Varid{for}\;{}\<[E]%
\\
\>[3]{}\hsindent{2}{}\<[5]%
\>[5]{}((\Varid{exampleTrace}\mathrel{\&\hspace{-2pt}\&\hspace{-2pt}\&\hspace{-2pt}\&}\Varid{haskellProgram})\mathbin{`\Varid{from}`}\Varid{randomSpecification})\;{}\<[E]%
\\
\>[3]{}\hsindent{2}{}\<[5]%
\>[5]{}\Varid{giveInteractionTrace}{}\<[E]%
\ColumnHook
\end{hscode}\resethooks
The names of the individual components and the usage of the combinators clearly communicate the basic idea of this \ensuremath{\Conid{TaskDesign}}.
The expression reads almost like actual instructions for a task.
This makes it easy for someone familiar with the EDSL, like a teaching assistant, to quickly modify and reuse parts of the design.

\section{Building Tasks on Haskell I/O}
\label{sec:examples}
With a general mechanism for describing tasks in place, we will now build some actual tasks on Haskell I/O programming.
The source code of the implementation and all examples from this paper, can be found at \url{https://github.com/fmidue/IOTasks}.%
\footnote{
The repository also contains instructions explaining how to generate and inspect random task instances, for the given examples.
}
First we introduce an alias for the type of inspectable I/O computations (see section~\ref{sec:specs}) to clearly separate it from syntactic program text.
\begin{hscode}\SaveRestoreHook
\column{B}{@{}>{\hspre}l<{\hspost}@{}}%
\column{E}{@{}>{\hspre}l<{\hspost}@{}}%
\>[B]{}\mathbf{type}\;\Conid{ExecutableHaskell}\mathrel{=}\mathit{IO}_\mathit{rep}\;(){}\<[E]%
\ColumnHook
\end{hscode}\resethooks
Our example tasks do not expect \ensuremath{\Conid{ExecutableHaskell}} as solution candidates but instead use syntactic \ensuremath{\Conid{HaskellCode}}.
For brevity we keep \ensuremath{\Conid{HaskellCode}} abstract here.
It is enough to know that we can inspect and print out values of this type.

Using \ensuremath{\Varid{fulfills}} from section~\ref{sec:specs} we can construct our first I/O specific requirement for a correct task solution.
\begin{hscode}\SaveRestoreHook
\column{B}{@{}>{\hspre}l<{\hspost}@{}}%
\column{E}{@{}>{\hspre}l<{\hspost}@{}}%
\>[B]{}\Varid{mustSatisfy}\mathbin{::}\Conid{Specification}\to \Conid{Require}\;\Conid{ExecutableHaskell}{}\<[E]%
\\
\>[B]{}\Varid{mustSatisfy}\;\Varid{s}\mathrel{=}\Varid{requireProp}\;(\mathbin{`\Varid{fulfills}`}\Varid{s}){}\<[E]%
\ColumnHook
\end{hscode}\resethooks
In order to be able to also check requirements like \ensuremath{\Varid{mustSatisfy}} we provide a function
\begin{hscode}\SaveRestoreHook
\column{B}{@{}>{\hspre}l<{\hspost}@{}}%
\column{E}{@{}>{\hspre}l<{\hspost}@{}}%
\>[B]{}\Varid{compile}\mathbin{::}\Conid{HaskellCode}\to \Conid{IO}\;(\Conid{Maybe}\;\Conid{ExecutableHaskell}){}\<[E]%
\ColumnHook
\end{hscode}\resethooks
to obtain semantic programs from syntactic representations.
Together with \ensuremath{\Varid{after}} we can now build semantic requirements for syntactic programs.
For example,
\begin{hscode}\SaveRestoreHook
\column{B}{@{}>{\hspre}l<{\hspost}@{}}%
\column{E}{@{}>{\hspre}l<{\hspost}@{}}%
\>[B]{}\Varid{passesCompiler}\mathbin{::}\Conid{Require}\;\Conid{HaskellCode}{}\<[E]%
\\
\>[B]{}\Varid{passesCompiler}\mathrel{=}\Varid{requirePure}\;(\Varid{const}\;\Conid{True})\mathbin{`\Varid{after}`}\Varid{compile}{}\<[E]%
\ColumnHook
\end{hscode}\resethooks
requires program text to be valid Haskell.
We will define additional requirements as we discuss the various example tasks.%
\footnote{All code shown in this section is part of the framework's API, except for expressions of type \ensuremath{\Varid{p}\to \Conid{TaskInstance}\;\Varid{s}} and \ensuremath{\Conid{TaskDesign}\;\Varid{s}}. Values of these two types should be read as defined by the framework's user.}

As hinted at earlier, we cannot rely only on verbal descriptions to convey a task's requirements.
Instead we will use fixed verbal instructions in combination with program code and/or interaction traces.
Our framework provides generators to build programs satisfying a given specification.
\begin{hscode}\SaveRestoreHook
\column{B}{@{}>{\hspre}l<{\hspost}@{}}%
\column{E}{@{}>{\hspre}l<{\hspost}@{}}%
\>[B]{}\Varid{haskellProgram}\mathbin{::}\Conid{Specification}\to \Conid{Gen}\;\Conid{HaskellCode}{}\<[E]%
\\
\>[B]{}\Varid{pythonProgram}\mathbin{::}\Conid{Specification}\to \Conid{Gen}\;\Conid{PythonCode}{}\<[E]%
\ColumnHook
\end{hscode}\resethooks
We use randomized generators to create program code from specifications as there usually are different ways to implement the given behavior.
Having access to different programs for the same specification is also useful for certain types of task designs, as we will see later on.
For program code we mainly use Haskell code in our tasks, but it is also useful to have access to code in other languages and paradigms.
For example, we use Python code to highlight how I/O looks different in Haskell compared to an imperative language.
Our students should already know Python as it is taught in their introductory programming course.

We will not discuss the full details of this code generation.
Our focus is on showcasing the different types of tasks expressible by the framework.

Essentially the code generation translates a given specification into an abstract program representation, agnostic about implementation techniques.
These programs can be translated to a concrete language's syntax, like Haskell or Python, by choosing appropriate embeddings of iteration, branching, state passing, etc.
Rewriting the intermediate representation, using predefined rules, we obtain slightly different programs for the same specification.
This approach also allows us to generate programs with (randomized) gaps or programs containing certain syntactic errors and anti-patterns.

The framework also provides generators for example traces matching a specification.
\begin{hscode}\SaveRestoreHook
\column{B}{@{}>{\hspre}l<{\hspost}@{}}%
\column{E}{@{}>{\hspre}l<{\hspost}@{}}%
\>[B]{}\Varid{exampleTrace}\mathbin{::}\Conid{Specification}\to \Conid{Gen}\;\Conid{Trace}{}\<[E]%
\\
\>[B]{}\Varid{exampleTraces}\mathbin{::}\Conid{Int}\to \Conid{Specification}\to \Conid{Gen}\;[\mskip1.5mu \Conid{Trace}\mskip1.5mu]{}\<[E]%
\ColumnHook
\end{hscode}\resethooks

Since all of the artifact generators require specifications as parameters, suitable generators are assumed to exist.
These generators are meant to be implemented by an educator/user as it is difficult to provide good generic generators.
A sketch of how to write generators for specifications can be found in \cite{flops20-paper}.
For our purposes we assume to have two generators:
\begin{hscode}\SaveRestoreHook
\column{B}{@{}>{\hspre}l<{\hspost}@{}}%
\column{E}{@{}>{\hspre}l<{\hspost}@{}}%
\>[B]{}\Varid{randomSpecification}\mathbin{::}\Conid{Gen}\;\Conid{Specification}{}\<[E]%
\\
\>[B]{}\Varid{similarSpecifications}\mathbin{::}\Conid{Gen}\;(\Conid{Specification},\Conid{Specification}){}\<[E]%
\ColumnHook
\end{hscode}\resethooks
The first generates a sensible random specification, for the educator's (task-specific) definition of sensible.
The second generates a pair of similar looking specifications with differing behavior.
Differing specification, for example, might have slightly different loop-termination conditions or varying outputs.

With all of these tools we now describe a diverse range of task designs.
Following Le~and~Pinkwart~\cite{le2014} we classify these tasks into three classes:
{
\renewcommand{\labelenumi}{\arabic{enumi})}
\begin{enumerate}
  \item Tasks with a single correct answer
  \item Tasks with multiple correct answers but only a single solution strategy
  \item Tasks with multiple different solution strategies
\end{enumerate}
}
In our case these classes correspond to the complexity of requirement descriptions and roughly to task difficulty.
Our introductory examples for programming tasks from section~\ref{sec:intro} sit at the two extreme points of this class spectrum.
The task with only a verbal description and no restrictions on the programming techniques to use is a perfect example of a class 3 task.
Adding additional requirements to such a task moves it further towards or into class 2.
On the other side of the spectrum, asking for a given program's behavior on a specific input is a class 1 task.
The rest of this section will explore different points on this spectrum.
Starting with class 1 tasks, we try to get as close to verbal-only free form tasks as possible.
Pedagogically, this spectrum can also be seen as a progression of consecutive exercise tasks, developing students abilities to read, reason about, expand and finally write programs~\cite{jeroen92}.

\subsection{Tasks with one correct answer}
Tasks with only a single correct answer cannot require a student to do any ``real'' programming.
Even for small programs there is almost never only one right answer.
Class 1 tasks are usually quiz-like tasks that focus on program reading and comprehension or simple program completion.

The simplest option to build such a task from our I/O related primitives is giving students two (or more) artifacts and asking them whether these artifacts originated from the same specification.
For example, given two programs, determine whether they have the same behavior.\\
\begin{hscode}\SaveRestoreHook
\column{B}{@{}>{\hspre}l<{\hspost}@{}}%
\column{3}{@{}>{\hspre}l<{\hspost}@{}}%
\column{5}{@{}>{\hspre}l<{\hspost}@{}}%
\column{E}{@{}>{\hspre}l<{\hspost}@{}}%
\>[B]{}\mathbf{data}\;\Conid{BinDesc}\mathrel{=}\Conid{Yes}\mid \Conid{No}\;\mathbf{deriving}\;(\Conid{Eq},\Conid{Ord},\Conid{Enum},\Conid{Show}){}\<[E]%
\\[\blanklineskip]%
\>[B]{}\Varid{decision}\mathbin{::}\Conid{TaskDesign}\;\Conid{BinDesc}{}\<[E]%
\\
\>[B]{}\Varid{decision}\mathrel{=}\Varid{for}\;(\Varid{equalityProblem}\mathbin{`\Varid{from}`}\Varid{similarSpecifications})\;\Varid{checkAgreement}{}\<[E]%
\\[\blanklineskip]%
\>[B]{}\Varid{checkAgreement}\mathbin{::}(\Conid{BinDesc},\Conid{HaskellCode},\Conid{HaskellCode})\to \Conid{TaskInstance}\;\Conid{BinDesc}{}\<[E]%
\\
\>[B]{}\Varid{checkAgreement}\;(\Varid{haveSameBehavior},p_1,p_2)\mathrel{=}\Conid{TaskInstance}{}\<[E]%
\\
\>[B]{}\hsindent{3}{}\<[3]%
\>[3]{}\{\mskip1.5mu \Varid{question}\mathrel{=}\Varid{text}\;\text{\ttfamily \char34 Do~these~two~programs~have~the~same~behavior?\char34}{}\<[E]%
\\
\>[3]{}\hsindent{2}{}\<[5]%
\>[5]{}\mathbin{\$\$}\Varid{text}\;(\Varid{show}\;p_1)\mathbin{\$\$}\Varid{text}\;\text{\ttfamily \char34 ---\char34}\mathbin{\$\$}\Varid{text}\;(\Varid{show}\;p_2){}\<[E]%
\\
\>[B]{}\hsindent{3}{}\<[3]%
\>[3]{},\Varid{given}\mathrel{=}\Conid{Nothing}{}\<[E]%
\\
\>[B]{}\hsindent{3}{}\<[3]%
\>[3]{},\Varid{requires}\mathrel{=}\Varid{exactAnswer}\;\Varid{haveSameBehavior}\mskip1.5mu\}{}\<[E]%
\ColumnHook
\end{hscode}\resethooks
For simplicity we assume that there are at least two different programs for each specification.
\begin{hscode}\SaveRestoreHook
\column{B}{@{}>{\hspre}l<{\hspost}@{}}%
\column{3}{@{}>{\hspre}l<{\hspost}@{}}%
\column{6}{@{}>{\hspre}l<{\hspost}@{}}%
\column{18}{@{}>{\hspre}l<{\hspost}@{}}%
\column{20}{@{}>{\hspre}l<{\hspost}@{}}%
\column{21}{@{}>{\hspre}l<{\hspost}@{}}%
\column{E}{@{}>{\hspre}l<{\hspost}@{}}%
\>[B]{}\Varid{equalityProblem}{}\<[18]%
\>[18]{}\mathbin{::}(\Conid{Specification},\Conid{Specification}){}\<[E]%
\\
\>[18]{}\hsindent{3}{}\<[21]%
\>[21]{}\to \Conid{Gen}\;(\Conid{BinDesc},\Conid{HaskellCode},\Conid{HaskellCode}){}\<[E]%
\\
\>[B]{}\Varid{equalityProblem}\;(s_1,s_2)\mathrel{=}\mathbf{do}{}\<[E]%
\\
\>[B]{}\hsindent{3}{}\<[3]%
\>[3]{}\Varid{sameBehavior}\leftarrow \Varid{elements}\;[\mskip1.5mu \Conid{No},\Conid{Yes}\mskip1.5mu]{}\<[E]%
\\
\>[B]{}\hsindent{3}{}\<[3]%
\>[3]{}(p_1,p_2)\leftarrow \mathbf{if}\;\Varid{sameBehavior}\mathrel{==}\Conid{Yes}{}\<[E]%
\\
\>[3]{}\hsindent{3}{}\<[6]%
\>[6]{}\mathbf{then}\;\Varid{differentPrograms}\;s_1\;s_1{}\<[E]%
\\
\>[3]{}\hsindent{3}{}\<[6]%
\>[6]{}\mathbf{else}\;\Varid{differentPrograms}\;s_1\;s_2{}\<[E]%
\\
\>[B]{}\hsindent{3}{}\<[3]%
\>[3]{}\Varid{pure}\;(\Varid{sameBehavior},p_1,p_2){}\<[E]%
\\[\blanklineskip]%
\>[B]{}\Varid{differentPrograms}{}\<[20]%
\>[20]{}\mathbin{::}\Conid{Specification}\to \Conid{Specification}{}\<[E]%
\\
\>[20]{}\to \Conid{Gen}\;(\Conid{HaskellCode},\Conid{HaskellCode}){}\<[E]%
\\
\>[B]{}\Varid{differentPrograms}\;s_1\;s_2\mathrel{=}\mathbf{do}{}\<[E]%
\\
\>[B]{}\hsindent{3}{}\<[3]%
\>[3]{}p_1\leftarrow \Varid{haskellProgram}\;s_1{}\<[E]%
\\
\>[B]{}\hsindent{3}{}\<[3]%
\>[3]{}p_2\leftarrow \Varid{haskellProgram}\;s_2\mathbin{`\Varid{suchThat}`}(\mathrel{/\hspace{-1ex}=}p_1){}\<[E]%
\\
\>[B]{}\hsindent{3}{}\<[3]%
\>[3]{}\Varid{return}\;(p_1,p_2){}\<[E]%
\ColumnHook
\end{hscode}\resethooks
To illustrate what instances of the \ensuremath{\Varid{decision}}-task from above can look like here is an example of two similar looking programs with slightly different behavior.
The second program is obtained by modifying the specification underlying the first program, in this case the loop-termination condition was randomly changed and the no longer needed initial input deleted.\\
\begin{minipage}{.5\textwidth}
\begin{hscode}\SaveRestoreHook
\column{B}{@{}>{\hspre}l<{\hspost}@{}}%
\column{3}{@{}>{\hspre}l<{\hspost}@{}}%
\column{9}{@{}>{\hspre}l<{\hspost}@{}}%
\column{11}{@{}>{\hspre}l<{\hspost}@{}}%
\column{13}{@{}>{\hspre}l<{\hspost}@{}}%
\column{E}{@{}>{\hspre}l<{\hspost}@{}}%
\>[B]{}p_1\mathrel{=}\mathbf{do}{}\<[E]%
\\
\>[B]{}\hsindent{3}{}\<[3]%
\>[3]{}\Varid{n}\leftarrow \Varid{readLn}{}\<[E]%
\\
\>[B]{}\hsindent{3}{}\<[3]%
\>[3]{}\mathbf{let}\;\Varid{loop}\;\Varid{xs}\mathrel{=}{}\<[E]%
\\
\>[3]{}\hsindent{6}{}\<[9]%
\>[9]{}\mathbf{if}\;\Varid{length}\;\Varid{xs}\mathrel{==}\Varid{n}{}\<[E]%
\\
\>[9]{}\hsindent{2}{}\<[11]%
\>[11]{}\mathbf{then}\;\mathbf{do}\;\Varid{return}\;\Varid{xs}{}\<[E]%
\\
\>[9]{}\hsindent{2}{}\<[11]%
\>[11]{}\mathbf{else}\;\mathbf{do}{}\<[E]%
\\
\>[11]{}\hsindent{2}{}\<[13]%
\>[13]{}\Varid{v}\leftarrow \Varid{readLn}{}\<[E]%
\\
\>[11]{}\hsindent{2}{}\<[13]%
\>[13]{}\Varid{loop}\;(\Varid{xs}\plus [\mskip1.5mu \Varid{v}\mskip1.5mu]){}\<[E]%
\\
\>[B]{}\hsindent{3}{}\<[3]%
\>[3]{}\Varid{ys}\leftarrow \Varid{loop}\;[\mskip1.5mu \mskip1.5mu]{}\<[E]%
\\
\>[B]{}\hsindent{3}{}\<[3]%
\>[3]{}\Varid{print}\;(\Varid{sum}\;\Varid{ys}){}\<[E]%
\ColumnHook
\end{hscode}\resethooks
\end{minipage}
\begin{minipage}{.5\textwidth}
\begin{hscode}\SaveRestoreHook
\column{B}{@{}>{\hspre}l<{\hspost}@{}}%
\column{3}{@{}>{\hspre}l<{\hspost}@{}}%
\column{9}{@{}>{\hspre}l<{\hspost}@{}}%
\column{11}{@{}>{\hspre}l<{\hspost}@{}}%
\column{13}{@{}>{\hspre}l<{\hspost}@{}}%
\column{E}{@{}>{\hspre}l<{\hspost}@{}}%
\>[B]{}p_2\mathrel{=}\mathbf{do}{}\<[E]%
\\
\>[B]{}\hsindent{3}{}\<[3]%
\>[3]{}\mathbf{let}\;\Varid{loop}\;\Varid{xs}\;\Varid{acc}\mathrel{=}{}\<[E]%
\\
\>[3]{}\hsindent{6}{}\<[9]%
\>[9]{}\mathbf{if}\;\Varid{acc}\mathrel{==}\mathrm{5}{}\<[E]%
\\
\>[9]{}\hsindent{2}{}\<[11]%
\>[11]{}\mathbf{then}\;\mathbf{do}\;\Varid{print}\;\Varid{xs}{}\<[E]%
\\
\>[9]{}\hsindent{2}{}\<[11]%
\>[11]{}\mathbf{else}\;\mathbf{do}{}\<[E]%
\\
\>[11]{}\hsindent{2}{}\<[13]%
\>[13]{}\Varid{v}\leftarrow \Varid{readLn}{}\<[E]%
\\
\>[11]{}\hsindent{2}{}\<[13]%
\>[13]{}\Varid{loop}\;(\Varid{xs}\mathbin{+}\Varid{v})\;(\Varid{acc}\mathbin{+}\mathrm{1}){}\<[E]%
\\
\>[B]{}\hsindent{3}{}\<[3]%
\>[3]{}\Varid{loop}\;\mathrm{0}\;\mathrm{0}{}\<[E]%
\ColumnHook
\end{hscode}\resethooks
\vspace{4ex}
\end{minipage}

Tasks on program completion use the possibility to generate partial programs we hinted at in section~\ref{sec:tasks}.
We use a generator
\begin{hscode}\SaveRestoreHook
\column{B}{@{}>{\hspre}l<{\hspost}@{}}%
\column{E}{@{}>{\hspre}l<{\hspost}@{}}%
\>[B]{}\Varid{haskellWithGaps}\mathbin{::}\Conid{Specification}\to \Conid{Gen}\;\Conid{HaskellCode}{}\<[E]%
\ColumnHook
\end{hscode}\resethooks
that produces an I/O program with gaps.
These gaps need to be filled with either \ensuremath{\Varid{readLn}} or \ensuremath{\Varid{print}}.
The different types of these two functions ensure there is only one correct solution.
For such a program we require choosing an appropriate expression for each gap.
\begin{hscode}\SaveRestoreHook
\column{B}{@{}>{\hspre}l<{\hspost}@{}}%
\column{3}{@{}>{\hspre}l<{\hspost}@{}}%
\column{7}{@{}>{\hspre}l<{\hspost}@{}}%
\column{E}{@{}>{\hspre}l<{\hspost}@{}}%
\>[B]{}\Varid{completion1}\mathbin{::}\Conid{TaskDesign}\;\Conid{HaskellCode}{}\<[E]%
\\
\>[B]{}\Varid{completion1}\mathrel{=}\Varid{for}\;(\Varid{haskellWithGaps}\mathbin{`\Varid{from}`}\Varid{randomSpecification})\;\Varid{fillGaps}{}\<[E]%
\\[\blanklineskip]%
\>[B]{}\Varid{fillGaps}\mathbin{::}\Conid{HaskellCode}\to \Conid{TaskInstance}\;\Conid{HaskellCode}{}\<[E]%
\\
\>[B]{}\Varid{fillGaps}\;\Varid{skeleton}\mathrel{=}\Conid{TaskInstance}{}\<[E]%
\\
\>[B]{}\hsindent{3}{}\<[3]%
\>[3]{}\{\mskip1.5mu \Varid{question}\mathrel{=}\Varid{text}\;\text{\ttfamily \char34 Complete~the~following~program.\char34}{}\<[E]%
\\
\>[3]{}\hsindent{4}{}\<[7]%
\>[7]{}\mathbin{\$\$}\Varid{text}\;\text{\ttfamily \char34 (Replace~?~with~readLn~or~print)\char34}{}\<[E]%
\\
\>[B]{}\hsindent{3}{}\<[3]%
\>[3]{},\Varid{given}\mathrel{=}\Conid{Just}\;\Varid{skeleton}{}\<[E]%
\\
\>[B]{}\hsindent{3}{}\<[3]%
\>[3]{},\Varid{requires}\mathrel{=}\Varid{passesCompiler}\mathbin{/\char92 }\Varid{mustMatch}\;\Varid{skeleton}\mskip1.5mu\}{}\<[E]%
\ColumnHook
\end{hscode}\resethooks

For the last type of answer, traces, we can give a Haskell program and some input sequence and ask students to execute the program on that input.
Fixing the input sequence, there is only one correct solution to such a task.
\begin{hscode}\SaveRestoreHook
\column{B}{@{}>{\hspre}l<{\hspost}@{}}%
\column{3}{@{}>{\hspre}l<{\hspost}@{}}%
\column{5}{@{}>{\hspre}l<{\hspost}@{}}%
\column{E}{@{}>{\hspre}l<{\hspost}@{}}%
\>[B]{}\mathit{comprehension}_1\mathbin{::}\Conid{TaskDesign}\;\Conid{Trace}{}\<[E]%
\\
\>[B]{}\mathit{comprehension}_1\mathrel{=}\Varid{for}{}\<[E]%
\\
\>[B]{}\hsindent{3}{}\<[3]%
\>[3]{}((\Varid{exampleTrace}\mathrel{\&\hspace{-2pt}\&\hspace{-2pt}\&\hspace{-2pt}\&}\Varid{haskellProgram})\mathbin{`\Varid{from}`}\Varid{randomSpecification}){}\<[E]%
\\
\>[B]{}\hsindent{3}{}\<[3]%
\>[3]{}\Varid{giveInteractionTrace}{}\<[E]%
\\[\blanklineskip]%
\>[B]{}\Varid{giveInteractionTrace}\mathbin{::}(\Conid{Trace},\Conid{HaskellCode})\to \Conid{TaskInstance}\;\Conid{Trace}{}\<[E]%
\\
\>[B]{}\Varid{giveInteractionTrace}\;(\Varid{t},\Varid{prog})\mathrel{=}\Conid{TaskInstance}{}\<[E]%
\\
\>[B]{}\hsindent{3}{}\<[3]%
\>[3]{}\{\mskip1.5mu \Varid{question}\mathrel{=}\Varid{text}\;(\text{\ttfamily \char34 Give~the~program's~trace~for~input~sequence\char34}){}\<[E]%
\\
\>[3]{}\hsindent{2}{}\<[5]%
\>[5]{}\mathbin{<>}\Varid{text}\;(\Varid{show}\;(\Varid{inputs}\;\Varid{t})){}\<[E]%
\\
\>[3]{}\hsindent{2}{}\<[5]%
\>[5]{}\mathbin{\$\$}\Varid{text}\;(\Varid{show}\;\Varid{prog}){}\<[E]%
\\
\>[B]{}\hsindent{3}{}\<[3]%
\>[3]{},\Varid{given}\mathrel{=}\Conid{Nothing}{}\<[E]%
\\
\>[B]{}\hsindent{3}{}\<[3]%
\>[3]{},\Varid{requires}\mathrel{=}\Varid{exactAnswer}\;\Varid{t}\mskip1.5mu\}{}\<[E]%
\\[\blanklineskip]%
\>[B]{}\Varid{inputs}\mathbin{::}\Conid{Trace}\to [\mskip1.5mu \Conid{String}\mskip1.5mu]{}\<[E]%
\ColumnHook
\end{hscode}\resethooks
This design can, for example, generate the task from page~\pageref{page:ex-task}.

\subsection{Tasks with multiple correct answers}
Before moving on to tasks on actual programming, we first look at a class 2 variant of the last task from the previous section.
Instead of giving a fixed input sequence we ask for an interaction trace with a certain property.
One possibility is to give two similar looking programs with different semantics and ask for an input sequence for which the given programs exhibit different I/O behavior.
\begin{hscode}\SaveRestoreHook
\column{B}{@{}>{\hspre}l<{\hspost}@{}}%
\column{3}{@{}>{\hspre}l<{\hspost}@{}}%
\column{5}{@{}>{\hspre}l<{\hspost}@{}}%
\column{19}{@{}>{\hspre}l<{\hspost}@{}}%
\column{E}{@{}>{\hspre}l<{\hspost}@{}}%
\>[B]{}\mathit{comprehension}_2\mathbin{::}\Conid{TaskDesign}\;[\mskip1.5mu \Conid{String}\mskip1.5mu]{}\<[E]%
\\
\>[B]{}\mathit{comprehension}_2\mathrel{=}\Varid{for}{}\<[E]%
\\
\>[B]{}\hsindent{3}{}\<[3]%
\>[3]{}((\Varid{specificationAnd}\;\Varid{haskellProgram}\mathrel{*\hspace{-3pt}*\hspace{-3pt}*\hspace{-1pt}*}\Varid{specificationAnd}\;\Varid{haskellProgram}){}\<[E]%
\\
\>[3]{}\hsindent{2}{}\<[5]%
\>[5]{}\mathbin{`\Varid{from}`}\Varid{similarSpecifications}){}\<[E]%
\\
\>[B]{}\hsindent{3}{}\<[3]%
\>[3]{}\Varid{findDiffSequence}{}\<[E]%
\\[\blanklineskip]%
\>[B]{}\Varid{findDiffSequence}{}\<[19]%
\>[19]{}\mathbin{::}((\Conid{Specification},\Conid{HaskellCode}),(\Conid{Specification},\Conid{HaskellCode})){}\<[E]%
\\
\>[19]{}\to \Conid{TaskInstance}\;[\mskip1.5mu \Conid{String}\mskip1.5mu]{}\<[E]%
\\
\>[B]{}\Varid{findDiffSequence}\;((s_1,p_1),(s_2,p_2))\mathrel{=}\Conid{TaskInstance}{}\<[E]%
\\
\>[B]{}\hsindent{3}{}\<[3]%
\>[3]{}\{\mskip1.5mu \Varid{question}\mathrel{=}\Varid{text}\;\text{\ttfamily \char34 Find~inputs~resulting~in~different~behavior.\char34}{}\<[E]%
\\
\>[3]{}\hsindent{2}{}\<[5]%
\>[5]{}\mathbin{\$\$}\Varid{text}\;(\Varid{show}\;p_1)\mathbin{\$\$}\Varid{text}\;\text{\ttfamily \char34 ---\char34}\mathbin{\$\$}\Varid{text}\;(\Varid{show}\;p_2){}\<[E]%
\\
\>[B]{}\hsindent{3}{}\<[3]%
\>[3]{},\Varid{given}\mathrel{=}\Conid{Nothing}{}\<[E]%
\\
\>[B]{}\hsindent{3}{}\<[3]%
\>[3]{},\Varid{requires}\mathrel{=}\Varid{triggerDifference}\;s_1\;s_2\mskip1.5mu\}{}\<[E]%
\\[\blanklineskip]%
\>[B]{}\Varid{triggerDifference}\mathbin{::}\Conid{Specification}\to \Conid{Specification}\to \Conid{Require}\;[\mskip1.5mu \Conid{String}\mskip1.5mu]{}\<[E]%
\\
\>[B]{}\Varid{triggerDifference}\;s_1\;s_2\mathrel{=}\Varid{requireProp}\mathbin{\$}\lambda \Varid{is}\to {}\<[E]%
\\
\>[B]{}\hsindent{3}{}\<[3]%
\>[3]{}((\mathrel{=\hspace{-1ex}/\hspace{-1ex}=})\mathbin{`\Varid{on}`}(\Varid{runProgram}\;\Varid{is}\mathbin{\circ}\Varid{buildComputation}))\;s_1\;s_2{}\<[E]%
\\[\blanklineskip]%
\>[B]{}\Varid{specificationAnd}{}\<[19]%
\>[19]{}\mathbin{::}(\Conid{Specification}\to \Conid{Gen}\;\Varid{a})\to \Conid{Specification}{}\<[E]%
\\
\>[19]{}\to \Conid{Gen}\;(\Conid{Specification},\Varid{a}){}\<[E]%
\\
\>[B]{}\Varid{specificationAnd}\;\Varid{g}\mathrel{=}\Varid{pure}\mathrel{\&\hspace{-2pt}\&\hspace{-2pt}\&\hspace{-2pt}\&}\Varid{g}{}\<[E]%
\ColumnHook
\end{hscode}\resethooks
The requirement uses the \ensuremath{\Varid{buildComputation}} function shown in section~\ref{sec:specs} to derive executable programs from the specifications.
Executing the specifications this way is easier than executing the displayed programs since these only exist in a textual form.

So far, our tasks are straightforward with regard to the questions asked.
For programming tasks beyond gap filling we now need to describe the required behavior of programs as well as restrict which solution strategies are valid.
To start off, we give interaction traces, i.e., example runs, to specify behavior and fix a solution strategy by providing a skeleton to  complete.
\begin{hscode}\SaveRestoreHook
\column{B}{@{}>{\hspre}l<{\hspost}@{}}%
\column{3}{@{}>{\hspre}l<{\hspost}@{}}%
\column{5}{@{}>{\hspre}l<{\hspost}@{}}%
\column{11}{@{}>{\hspre}l<{\hspost}@{}}%
\column{E}{@{}>{\hspre}l<{\hspost}@{}}%
\>[B]{}\mathit{completion}_2\mathbin{::}\Conid{TaskDesign}\;\Conid{HaskellCode}{}\<[E]%
\\
\>[B]{}\mathit{completion}_2\mathrel{=}\Varid{for}{}\<[E]%
\\
\>[B]{}\hsindent{3}{}\<[3]%
\>[3]{}(\Varid{exampleTraces}\;\mathrm{5}\mathbin{`\Varid{from}`}\Varid{fixed}\;\Varid{specification}){}\<[E]%
\\
\>[B]{}\hsindent{3}{}\<[3]%
\>[3]{}\Varid{matchExamples}{}\<[E]%
\\
\>[B]{}\hsindent{3}{}\<[3]%
\>[3]{}\mathbf{where}\;\Varid{specification}\mathrel{=}\dots{}\<[E]%
\\[\blanklineskip]%
\>[B]{}\Varid{matchExamples}\mathbin{::}[\mskip1.5mu \Conid{Trace}\mskip1.5mu]\to \Conid{TaskInstance}\;\Conid{HaskellCode}{}\<[E]%
\\
\>[B]{}\Varid{matchExamples}\;\Varid{ts}\mathrel{=}\Conid{TaskInstance}{}\<[E]%
\\
\>[B]{}\hsindent{3}{}\<[3]%
\>[3]{}\{\mskip1.5mu \Varid{question}\mathrel{=}\Varid{text}\;\text{\ttfamily \char34 Complete~the~program~to~match~the~examples:\char34}{}\<[E]%
\\
\>[3]{}\hsindent{2}{}\<[5]%
\>[5]{}\mathbin{\$\$}\Varid{vcat}\;(\Varid{map}\;(\Varid{text}\mathbin{\circ}\Varid{show})\;\Varid{ts}){}\<[E]%
\\
\>[B]{}\hsindent{3}{}\<[3]%
\>[3]{},\Varid{given}\mathrel{=}\Conid{Just}\;\Varid{skeleton}{}\<[E]%
\\
\>[B]{}\hsindent{3}{}\<[3]%
\>[3]{},\Varid{requires}\mathrel{=}\Varid{produceTraces}\;\Varid{ts}\mathbin{`\Varid{after}`}\Varid{compile}\mathbin{/\char92 }\Varid{mustMatch}\;\Varid{skeleton}\mskip1.5mu\}{}\<[E]%
\\
\>[B]{}\hsindent{3}{}\<[3]%
\>[3]{}\mathbf{where}\;\Varid{skeleton}\mathrel{=}\Varid{fromSourceString}\mathbin{\$}\Varid{unlines}{}\<[E]%
\\
\>[3]{}\hsindent{8}{}\<[11]%
\>[11]{}[\mskip1.5mu \text{\ttfamily \char34 main~=~do\char34}{}\<[E]%
\\
\>[3]{}\hsindent{8}{}\<[11]%
\>[11]{},\text{\ttfamily \char34 ~~?\char34}{}\<[E]%
\\
\>[3]{}\hsindent{8}{}\<[11]%
\>[11]{},\text{\ttfamily \char34 ~~while~?~?~?\char34}{}\<[E]%
\\
\>[3]{}\hsindent{8}{}\<[11]%
\>[11]{},\text{\ttfamily \char34 while~::~(a~->~Bool)~->~(a~->~IO~a)~->~a~->~IO~a\char34}{}\<[E]%
\\
\>[3]{}\hsindent{8}{}\<[11]%
\>[11]{},\text{\ttfamily \char34 while~=~...\char34}\mskip1.5mu]{}\<[E]%
\\[\blanklineskip]%
\>[B]{}\Varid{produceTraces}\mathbin{::}[\mskip1.5mu \Conid{Trace}\mskip1.5mu]\to \Conid{Require}\;\Conid{ExecutableHaskell}{}\<[E]%
\\
\>[B]{}\Varid{produceTraces}\;\Varid{ts}\mathrel{=}\Varid{requirePure}\mathbin{\$}\lambda \Varid{p}\to {}\<[E]%
\\
\>[B]{}\hsindent{3}{}\<[3]%
\>[3]{}\Varid{all}\;(\lambda \Varid{t}\to \Varid{runProgram}\;(\Varid{inputs}\;\Varid{t})\;\Varid{p}\mathrel{==}\Varid{t})\;\Varid{ts}{}\<[E]%
\\[\blanklineskip]%
\>[B]{}\Varid{fromSourceString}\mathbin{::}\Conid{String}\to \Conid{HaskellCode}{}\<[E]%
\ColumnHook
\end{hscode}\resethooks
This task asks to produce the given examples by mimicking an imperative loop using the higher-order \ensuremath{\Varid{while}} function.
For simplicity we use a fixed specification instead of a random one.
Using a random specification, without any further restrictions, potentially results in an unsolvable task since the skeleton can have the wrong structure for the underlying behavior.
With a suitable generator, however, the underlying specification can be randomized as well.

Using traces to describe behavior requirements has a disadvantage one needs to be aware of.
Such descriptions may not fully characterize the underlying specification's behavior.
The description essentially gives a list of unit tests for a solution to fulfill.
The underlying specification guarantees there is at least one program solving the task without hard-coding the given examples.
However, a program working only for the given examples and crashing on all others inputs still meets the requirement.
In the above task such solutions are largely ruled out by the skeleton.
For free-form programming tasks that do not use a skeleton, this is a more serious problem.
Our solution is to use program code itself to describe the required behavior.
Doing so naturally gives rise to two exercise types: refactoring and cross-language re-implementation.

Refactoring tasks ask to rewrite a given program into a program that satisfies certain properties the original program does not have.
Both programs should behave identically, with regards to I/O, for the same inputs.
The original program therefore fully describes the behavior a correct solution should have.

The next example gives a program accumulating list values and outputting the result of a computation, expressible as a fold, on this list.
Students are asked to rewrite this program into a version not using any list, directly carrying out the computation.
\begin{hscode}\SaveRestoreHook
\column{B}{@{}>{\hspre}l<{\hspost}@{}}%
\column{3}{@{}>{\hspre}l<{\hspost}@{}}%
\column{7}{@{}>{\hspre}l<{\hspost}@{}}%
\column{E}{@{}>{\hspre}l<{\hspost}@{}}%
\>[B]{}\Varid{refactoring}\mathbin{::}\Conid{TaskDesign}\;\Conid{HaskellCode}{}\<[E]%
\\
\>[B]{}\Varid{refactoring}\mathrel{=}\Varid{for}{}\<[E]%
\\
\>[B]{}\hsindent{3}{}\<[3]%
\>[3]{}(\Varid{specificationAnd}\;\Varid{haskellFoldProgram}\mathbin{`\Varid{from}`}\Varid{fixed}\;\Varid{specification}){}\<[E]%
\\
\>[B]{}\hsindent{3}{}\<[3]%
\>[3]{}\Varid{rewriteToNoLists}{}\<[E]%
\\
\>[B]{}\hsindent{3}{}\<[3]%
\>[3]{}\mathbf{where}\;\Varid{specification}\mathrel{=}\mathit{undefined}{}\<[E]%
\\[\blanklineskip]%
\>[B]{}\Varid{rewriteToNoLists}\mathbin{::}(\Conid{Specification},\Conid{Description})\to \Conid{TaskInstance}\;\Conid{HaskellCode}{}\<[E]%
\\
\>[B]{}\Varid{rewriteToNoLists}\;(\Varid{spec},\Varid{prog})\mathrel{=}\Conid{TaskInstance}{}\<[E]%
\\
\>[B]{}\hsindent{3}{}\<[3]%
\>[3]{}\{\mskip1.5mu \Varid{question}\mathrel{=}{}\<[E]%
\\
\>[3]{}\hsindent{4}{}\<[7]%
\>[7]{}\Varid{text}\;\text{\ttfamily \char34 Re-write~the~program~such~that~it~does~not~use~lists.\char34}{}\<[E]%
\\
\>[3]{}\hsindent{4}{}\<[7]%
\>[7]{}\mathbin{\$\$}\Varid{prog}{}\<[E]%
\\
\>[B]{}\hsindent{3}{}\<[3]%
\>[3]{},\Varid{given}\mathrel{=}\Conid{Nothing}{}\<[E]%
\\[\blanklineskip]%
\>[B]{}\hsindent{3}{}\<[3]%
\>[3]{},\Varid{requires}\mathrel{=}(\Varid{mustSatisfy}\;\Varid{spec}\mathbin{`\Varid{after}`}\Varid{compile})\mathbin{/\char92 }\Varid{noLists}\mskip1.5mu\}{}\<[E]%
\\[\blanklineskip]%
\>[B]{}\Varid{noLists}\mathbin{::}\Conid{Require}\;\Conid{HaskellCode}{}\<[E]%
\\
\>[B]{}\Varid{noLists}\mathrel{=}\Varid{requirePure}\mathbin{\$}\lambda \Varid{p}\to {}\<[E]%
\\
\>[B]{}\hsindent{3}{}\<[3]%
\>[3]{}not\;(\Varid{containsFunction}\;\text{\ttfamily \char34 ++\char34}\;\Varid{p}\mathrel{\vee}\Varid{containsFunction}\;\text{\ttfamily \char34 :\char34}\;\Varid{p}){}\<[E]%
\\[\blanklineskip]%
\>[B]{}\Varid{containsFunction}\mathbin{::}\Conid{String}\to \Conid{HaskellCode}\to \Conid{Bool}{}\<[E]%
\\[\blanklineskip]%
\>[B]{}\Varid{haskellFoldProgram}\mathbin{::}\Conid{Specification}\to \Conid{Gen}\;\Conid{Description}{}\<[E]%
\ColumnHook
\end{hscode}\resethooks
We check the requirement of not using lists by simply verifying that solution code does not contain the functions to build lists.
Once again, the exact details on how \ensuremath{\Varid{haskellFoldProgram}} is implemented internally are outside the scope of this presentation.
We basically take a program and look for a function known to be a fold, e.g. \ensuremath{\Varid{sum}} or \ensuremath{\Varid{length}}, that is used on the result of a list accumulating loop.
This function's base case and recursive step are then ``inlined'' into the loop.

\subsection{Tasks with different solution strategies}
For our last example we finally arrive at a task very similar to the verbal-only free-form tasks from the beginning.
The task requires re-implementing a Python program in Haskell.
As already stated above, we choose Python since our students learn it as their first programming language and are therefore already familiar with it.
\begin{hscode}\SaveRestoreHook
\column{B}{@{}>{\hspre}l<{\hspost}@{}}%
\column{3}{@{}>{\hspre}l<{\hspost}@{}}%
\column{5}{@{}>{\hspre}l<{\hspost}@{}}%
\column{E}{@{}>{\hspre}l<{\hspost}@{}}%
\>[B]{}\Varid{pythonToHaskell}\mathbin{::}\Conid{TaskDesign}\;\Conid{HaskellCode}{}\<[E]%
\\
\>[B]{}\Varid{pythonToHaskell}\mathrel{=}\Varid{for}{}\<[E]%
\\
\>[B]{}\hsindent{3}{}\<[3]%
\>[3]{}(\Varid{specificationAnd}\;\Varid{pythonProgram}\mathbin{`\Varid{from}`}\Varid{randomSpecification}){}\<[E]%
\\
\>[B]{}\hsindent{3}{}\<[3]%
\>[3]{}\Varid{rewriteAsHaskell}{}\<[E]%
\\[\blanklineskip]%
\>[B]{}\Varid{rewriteAsHaskell}\mathbin{::}(\Conid{Specification},\Conid{PythonCode})\to \Conid{TaskInstance}\;\Conid{HaskellCode}{}\<[E]%
\\
\>[B]{}\Varid{rewriteAsHaskell}\;(\Varid{s},\Varid{prog})\mathrel{=}\Conid{TaskInstance}{}\<[E]%
\\
\>[B]{}\hsindent{3}{}\<[3]%
\>[3]{}\{\mskip1.5mu \Varid{question}\mathrel{=}\Varid{text}\;\text{\ttfamily \char34 Write~the~following~program~in~Haskell:\char34}{}\<[E]%
\\
\>[3]{}\hsindent{2}{}\<[5]%
\>[5]{}\mathbin{\$\$}\Varid{text}\;(\Varid{show}\;\Varid{prog}){}\<[E]%
\\
\>[B]{}\hsindent{3}{}\<[3]%
\>[3]{},\Varid{given}\mathrel{=}\Conid{Nothing}{}\<[E]%
\\
\>[B]{}\hsindent{3}{}\<[3]%
\>[3]{},\Varid{requires}\mathrel{=}\Varid{mustSatisfy}\;\Varid{s}\mathbin{`\Varid{after}`}\Varid{compile}\mskip1.5mu\}{}\<[E]%
\ColumnHook
\end{hscode}\resethooks
This task does not fix a solution strategy.
By starting from a Python program it precisely states the required behavior.
The Python program also does not contain any information on how a Haskell program with the identical behavior could look like.
I/O programs in Haskell usually have a different structure compared to imperative languages (e.g. recursive functions vs. explicit loops).
Take, for example, the following task instance generated from the above design.
{\small
\begin{tabbing}\ttfamily
~Re\char45{}implement~the~following~Python~program~in~Haskell\char58{}\\
\ttfamily ~n~\char61{}~int\char40{}input\char40{}\char41{}\char41{}\\
\ttfamily ~x~\char61{}~\char91{}\char93{}\\
\ttfamily ~while~len\char40{}x\char41{}~\char33{}\char61{}~n~\char58{}\\
\ttfamily ~~~v~\char61{}~int\char40{}input\char40{}\char41{}\char41{}\\
\ttfamily ~~~x~\char43{}\char61{}~\char91{}v\char93{}\\
\ttfamily ~print\char40{}sum\char40{}x\char41{}\char41{}
\end{tabbing}
}
The required behavior is the same as the verbal description from the introduction.
It is clear that we need to use some form of repetition in our solution, but that information is also contained in the verbal description (``\textit{read $n$ integers one after the other}'').
Only the usage of a list to store read integers is not in the verbal description.
For everyone with basic programming skills this should not be anything new.
We therefore argue that the above Python program can~replace the verbal description without loosing precision or providing additional hints.

\section{Related Work}
Tools for automatic task generation exist in a variety of different application areas, for example, general science questions~\cite{DBLP:conf/aclnut/WelblLG17}, math related tasks~\cite{DBLP:conf/latice/Kurt-KaraogluSS15} and programming tasks~\cite{DBLP:conf/norchip/MosbeckHJ18,DBLP:conf/csedu/StrieweBG09}.
Some systems for natural language questions can generate tasks from databases of domain specific text~\cite{DBLP:conf/aclnut/WelblLG17}, but most approaches use templates together with parameter generators, similar to our \ensuremath{\Conid{TaskDesign}}s.
In contrast to our flexible EDSL approach most of these systems use rigid template formats provided as inputs to the task generation.
Consequently, these systems are usually embedded inside a specific e-learning environment.
Our framework can, in principle, be used with any e-learning system.
We already use a modular e-learning system \cite{rahn2008leipzig,DBLP:conf/abp/Waldmann17} and plan to integrate the framework in that context.

Our generated task instances do not provide any detailed feedback apart from maybe some QuickCheck outputs.
There are other automatic assessment systems providing more detailed feedback, including suggestions on how to fix mistakes.
A survey of different automatic assessment systems for programming tasks and the feedback they can generate is presented by Keuning et al.~\cite{keuning2019}.

\section{Conclusion \& Future work}
The presented framework can be used to describe a diverse range of exercise task designs and generate concrete randomized instances from these designs.
Separating task descriptions into orthogonal components makes modifying and reusing tasks easy.
Task designs also serve as high-level documentation for the task's idea if we choose descriptive names for the individual components.

We presented examples of using the framework to describe tasks on Haskell I/O.
The variety in these tasks stems from domain specific primitives providing different artifacts around which the parameterized tasks are built.
These artifacts are used as stand-ins for verbal descriptions to precisely state task requirements even if tasks are built from randomized specifications.
Tasks created this way have a slightly different feel compared to traditional hand-written ones.
Even though there are some restrictions to our approach, we can still create a wide range of different task types.
To the best of our knowledge automatically deriving artifacts for communicating requirements is a novel approach to automatic task generation in the context of programming tasks.

We have not yet had the opportunity to use tasks like the ones shown in section~\ref{sec:examples} in practice. %
However, our tasks on Haskell I/O-programming already use the specification language and its testing facilities.
We plan to test the presented approach to task design in the next iteration of our programming paradigms course.
We are especially interested to see whether tasks on program completion and comprehension benefit students when learning Haskell-I/O.
Writing high-quality generators for specifications should also be investigated further.
Good generators have a big influence on the quality of concrete task instances.

\bibliographystyle{splncs04}
\bibliography{references}
\end{document}